# Lithium Metal Penetration Induced by Electrodeposition through Solid Electrolytes: Example in Single-Crystal $Li_6La_3ZrTaO_{12}$ Garnet

Tushar Swamy[1], Richard Park[2], Brian W. Sheldon[3], Daniel Rettenwander[4], Lukas Porz[5], Stefan Berendts[6], Reinhard Uecker[7], W. Craig Carter[2] and Yet-Ming Chiang[2, *]

[1]Department of Mechanical Engineering, [2]Department of Material Science & Engineering
Massachusetts Institute of Technology, Cambridge, 02139 USA

[3]School of Engineering, Brown University, Providence, Rhode Island 02019, United States

[4]Institute for Chemistry and Technology of Materials, Graz University of Technology, 8010 Graz, Austria

[5]Institute of Material Science, Technische Universität Darmstadt, 64287 Darmstadt, Germany

[6]Department of Chemistry, Technische Universität Berlin, 10623 Berlin, Germany

[7]Leibniz Institute for Crystal Growth, 12489 Berlin, Germany

[*]Corresponding author: ychiang@mit.edu

**Abstract:**

Solid electrolytes are considered a potentially enabling component in rechargeable batteries that use lithium metal as the negative electrode, and thereby can safely access higher energy density than available with today's lithium ion batteries. To do so, the solid electrolyte must be able to suppress morphological instabilities that lead to poor coulombic efficiency and, in the worst case, internal short circuits. In this work, lithium electrodeposition experiments were performed using single-crystal $Li_6La_3ZrTaO_{12}$ garnet as solid electrolyte layers to investigate the factors that determine whether lithium penetration occurs through brittle inorganic solid electrolytes. In these single crystals, grain boundaries are excluded as possible paths for lithium metal propagation. However, Vickers microindentation was used to introduce sharp surface flaws of known size. Using *operando* optical microscopy, it was found that lithium metal penetration sometimes initiates at these controlled surface defects, and when multiple indents of varying size were present, propagates preferentially from the largest defect. However, a second class of flaws was found to be equally or more important. At the perimeter of surface current collectors, an enhanced electrodeposition current density causes lithium metal filled cracks to initiate and grow to penetration, even when the large Vickers defects are in close proximity. Modeling the electric field concentration for the experimental configurations, it was shown that a factor of 5 enhancement in field can readily occur within 10 micrometers of current collector discontinuities, which we interpret as the origin of electrochemomechanical stresses leading to failure. Such field amplification may determine the sites where supercritical surface defects dominate lithium metal propagation during electrodeposition, overriding the presence of larger defects elsewhere.

**Broader Context**

All-solid-state batteries can potentially store electricity at higher energy density and with greater safety than existing lithium-ion technology but require the use of lithium metal electrodes. Towards these goals, it is critical to understand possible failure modes when lithium metal electrodes are used with solid electrolytes, and especially the processes of metal dendrite formation and propagation. Here, we test the stability limits of lithium metal electrodeposition using high quality single crystals of LLZTO garnet solid electrolyte, at high current densities (5 to 10 mA/cm$^2$) equivalent to charging a battery at 1C-2C rates (1h to 0.5h charge times). We surprisingly observe that lithium metal filled cracks initiate at the edges of surface metal current collectors, rather than on millimeter-scale deliberately introduced surface cracks. At these current densities, lithium metal penetrates to short-circuit through ~2mm electrolyte thickness on the minute time scale. The results highlight a previously unrecognized failure mode for all-solid-state batteries and suggest that control of electric field distributions will be critical to successful implementation.

## 1. Introduction

Rechargeable batteries have enabled multiple advances in portable electronics, transportation, and renewable energy storage. For example, today's electric vehicle (EV) grade lithium-ion batteries possess an industry leading combination of high specific energy (~150 Wh·kg$^{-1}$, pack-level) and energy density (~250 Wh·L$^{-1}$, pack-level).[1] However, to meet future energy and cost targets that at least double the energy density,[2] radical improvements are necessary. One promising approach, common to all of the "beyond lithium ion" approaches currently cited by the U.S. Department of Energy's Vehicle Technologies office,[3] is to use a metallic negative electrode, especially lithium,[4–6] which is the most electropositive (-3.04 V w.r.t the standard hydrogen electrode) and lightest metal (density of 0.534 g·cm$^{-3}$), resulting in an exceptionally high capacity (3869 mAh·g$^{-1}$, 2066 Ah·L$^{-1}$).

Although using lithium metal anodes with conventional liquid electrolytes would achieve target energy densities, there have been significant historical challenges in enabling such a battery. These include the formation of an unstable solid-electrolyte interphase on the lithium metal that continuously depletes the electrolyte during cycling, and a susceptibility to short circuits due to dendrite formation through the liquid electrolyte at practical current densities (1-3 mA·cm$^{-2}$). [7–9]

An alternative approach to enabling lithium metal anodes is to pair them with ceramic electrolytes, which could potentially result in safe, energy-dense batteries surpassing state-of-the-art lithium-ion batteries (~150 Wh·kg$^{-1}$ to ~250 Wh·kg$^{-1}$ / ~250 Wh·L$^{-1}$ to ~750 Wh·L$^{-1}$). [4,10–14] Inorganic solid-state electrolytes (SSEs) such as lithium sulfides (*e.g.* β-Li$_3$PS$_4$) and garnet-structure oxides (*e.g.* Li$_7$La$_3$Zr$_2$O$_{12}$) have received much attention in recent years due to their high ionic conductivity,[15,16] and progress towards achieving an electrochemically stable electrode-electrolyte interface.[11,14] Furthermore, inorganic solid electrolytes possess high elastic moduli, which has been identified as one criterion necessary for suppression of lithium metal dendrites.[17] Nonetheless, multiple groups, have reported formation of short circuits when polycrystalline SSEs are cycled in contact with a lithium metal anode above a critical current density.[18–22]

To explain this discrepancy, our group previously performed cyclic electrodeposition experiments using symmetric Li/SSE/Li cells in which the SSE is amorphous, single crystalline, or polycrystalline.[23] lithium metal penetration leading to electrical shorting was observed in all three classes of SSEs, including for a garnet oxide having a shear modulus far above the criterion suggested by Monroe and Newman[17] (e.g., at least twice that of lithium metal). In the polycrystalline SSEs, lithium propagation appeared to preferentially follow grain boundaries, which others have also observed. [18] However, even in amorphous LPS and single crystalline LLZTO garnet, propagation of lithium metal-filled cracks was observed. These results point to a mechanism for lithium penetration whereby the lithium "dendrites" initiate at surface flaws. Results suggested that when lithium metal electroplates onto the SSE, pre-existing surface flaws, which may include grain boundary grooves or multigrain junctions if the SSE is polycrystalline, are first filled with electroplated metal. These become preferred sites for subsequent electrodeposition, thereby driving fracture of the bulk electrolyte via stress amplification at crack tips, analogous to established models for brittle fracture under purely mechanical loads. [24-27]

In this previous work, we proposed that short circuits in single crystal electrolytes occur as a result of stress generation sufficient to fracture the electrolyte during metal electrodeposition. [23] Two mutually compatible processes that dictate stress generation are considered in this work. First, a surface (or internal) flaw leads to current focusing that locally enhances the lithium flux. Lithium accumulation inside of this flaw can generate internal stress, which is thermodynamically limited by the electrical overpotential (i.e., if the stress is sufficiently high it will counteract the electrochemical driving force). This internal stress causes extension of the Li-filled surface flaw if the elastic energy release rate exceeds the fracture resistance of the electrolyte. [23] The second process that must be considered is that removal of lithium from the flaw will mitigate the internal stress build-up. For example, prior work with liquid sodium electrodes considers analogous models based on Poiseuille flow, which is valid for laminar flow of liquids. In our case lithium "extrusion" out of filled flaws may occur due to creep and/or plastic deformation of the metal. In lithium, these mechanisms are currently not well understood, particularly at the small length scales that are relevant inside of solid electrolytes. [28] However, the balance between the lithium insertion and removal mechanisms that are outlined above produces a unique relationship between the electrodeposition current density and the stress required to extend metal filled cracks. [29–38]

It should be noted that the equilibrium hydrostatic stresses which correspond to the applied overpotential are a thermodynamic upper bound on the electrochemically generated stress inside of the flaw. Stress relaxation due to extrusion (or possibly other processes) can then lead to a lower, kinetically limited stresses inside of the actual flaw.

Based on this interplay between lithium insertion and removal, the threshold for the extension of a lithium filled filament through a solid electrolyte can be described in terms of either a critical overpotential or current density. That is, at any given overpotential (or current density), there is a critical flaw size above which extension is energetically favorable. The relationship between this critical size and overpotential (or current density) is in principle amenable to experimental evaluation. While the overall electrical potential or current density is easily controlled in typical electrochemical cells, it is difficult to know the distribution of defects in the system, and especially at the SSE surface. The basic energetic argument that we have put forth is analogous to Griffith theory, wherein the largest defect is expected to propagate most readily (i.e., at the lowest critical overpotential or critical current density). Thus, we conducted the converse experiment to test our hypothesis: Sharp defects of such large size that they assuredly lie at the high end of the flaw size distribution were introduced through microindentation (here with a Vickers square pyramid). These were then used to investigate lithium metal propagation as a function of overpotential and current density. However, in the course of investigating the effects of large controlled flaws, we discovered that variations in current density due to the geometric configuration of surface current collectors can have a larger, dominant effect on lithium penetration.

## 2. Experimental Design

Our experimental configuration creates a lithium flux from a lithium metal counterelectrode (CE) through a highly polished single crystal of $Li_6La_3ZrTaO_{12}$ (sc-LLZTO) garnet to a gold working electrode (WE), at which the lithium metal electrodeposits. *Operando* observations were made (**Figure 1**) of the morphology of the deposited lithium metal, and the initiation and growth of lithium metal filled cracks (if present) at the WE/SSE interface. Single crystal LLZTO was used to create a "best case" situation where the grain boundaries and residual porosity present in polycrystals are excluded as possible initiation sites. A tungsten needle was used to make electrical contact to the surface of the gold working electrode. Galvanostatic experiments were conducted, at current densities producing a cell voltage of 3.5 ~ 4V, which based on a previous failure model, should be sufficient to propagate even nm-scale surface flaws.[23] Optical microscopy was used to observe the lithium plating behavior in real time, during which the entire apparatus was housed in an argon-filled glove box with oxygen and water content below 0.1 ppm and 1 ppm, respectively. The ionic conductivity of the sc-LLZTO SSE was previously measured via electrochemical impedance spectroscopy (EIS) to be 0.2 mS·cm$^{-1}$.[23] The crystal structure, single crystallinity, and phase purity of the sc-LLZTO have previously been confirmed via X-ray diffraction (XRD).[39]

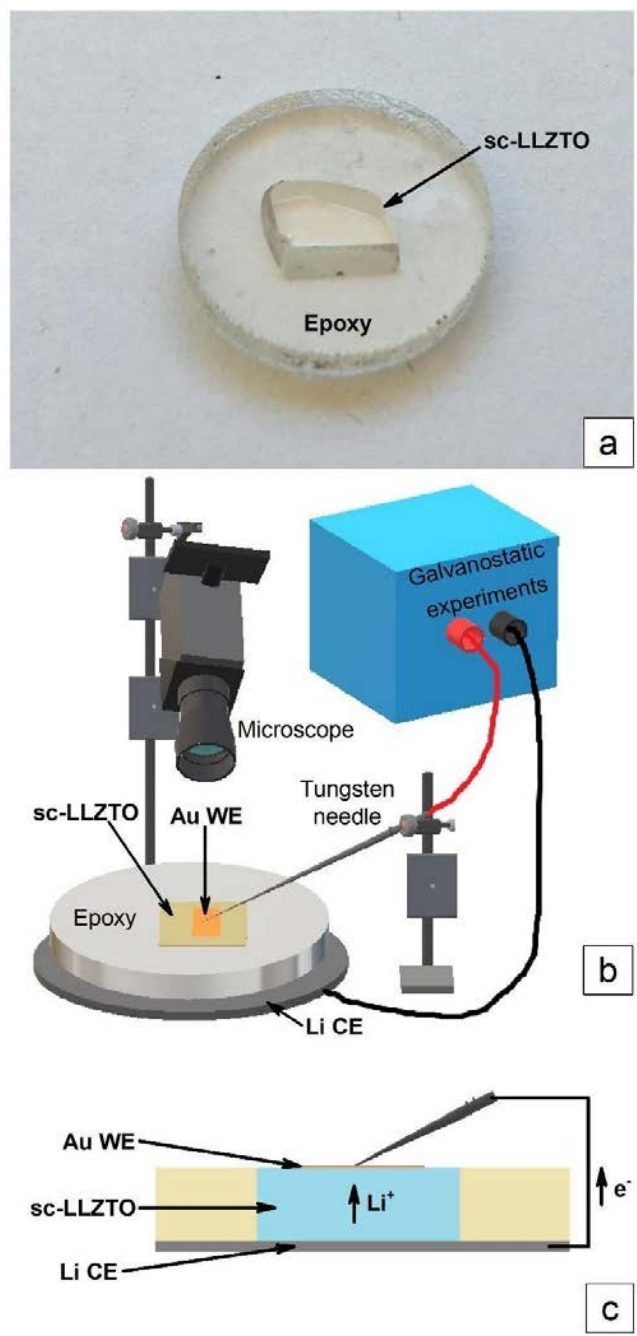

**Figure 1.** (a) Example of single-crystal electrolyte samples used in this study, polished and embedded in epoxy for handling. (b) Schematic of the apparatus used for galvanostatic lithium plating from a lithium metal counterelectrode (CE) through the sc-LLZTO onto a gold working electrode (WE). (c) Cross sectional schematic of the electrodeposition experiment.

## 1.1 Preparation of Single Crystal LLZTO Samples with Controlled Surface Flaws

Slices of sc-LLZTO having typical dimensions of 5 x 5 x 2 mm were embedded in epoxy for ease of handling. The two largest parallel surfaces were polished with diamond abrasive paste of successively finer grit size, to a final grit size of 1 μm resulting in a mirror finish. The relatively large thicknesses were used to facilitate observations of flaw propagation at optical microscopy resolution. Typical scanning electron microscopy (SEM) and atomic force microscopy (AFM) images of the polished surface are shown in **Figures 2(a)** and **2(b)**, respectively. The SEM image is nearly featureless, whereas the AFM scan reveals linear features of sub-micrometer width resulting from polishing. Still, the polished surfaces have a rms roughness of only 0.2 nm (measured over a 10 μm x 10 μm scan area). As previously noted, [23] even highly polished surfaces will have a distribution of defect sizes, amongst which the largest, controlling flaws can be exceedingly difficult to find. Therefore, we elected to introduce surface flaws of controlled and reproducible size that are much larger than any resulting from polishing, to establish the largest, and theoretically the controlling, flaw size. A Vickers micro-indenter was used to place several (typically four) indents under an applied load sufficient to produce cracks that emanate from the corners. These cracks have a semicircular crack front that extends normal to the surface into the crystal. By varying the applied load, a range of crack lengths is produced. An optical microscopy image of a typical indent array, and an SEM image of an indent showing the characteristic corner cracks, appear in **Figure 2(c)** and **2(d)**, respectively. For indentation loads of 2N, 5N, 10N, and 20N, the crack lengths were measured via optical microscopy to be about 50, 140, 270, and 400 μm, respectively. The corresponding hardness (determined from the square-pyramid indent size) and fracture toughness (determined from the crack length) of the LLZTO were calculated to be 7.5 GPa and 0.6 MPa·m$^{1/2}$, respectively[40], assuming a Young's modulus of 150 GPa as measured previously.[36] These values are reasonably close to Wolfenstine *et al.*'s results for 97% dense polycrystalline $Li_{6.28}La_3Zr_2Al_{0.24}O_{12}$ (LLZAO) garnet, 6.3 GPa and 1.14 MPa·m$^{1/2}$, respectively.[41] Our slightly higher hardness and slightly lower fracture toughness are reasonable for a single crystal compared to a polycrystal of the same basic composition, as a polycrystal can be somewhat softer due to residual porosity, but also somewhat tougher as the microstructure can deflect cracks. These values also indicate that LLZO-based garnets, as a family, are both softer and less tough than most oxide ceramics. Polycrystalline ceramics typically have fracture toughnesses in the 2-4.5 MPa·m$^{1/2}$ range. LLZTO has a low single crystal fracture toughness close to that of ordinary soda-lime-silicate glass (~0.7 MPa·m$^{1/2}$),[42] although the $Li_2S_2$-$P_2S_5$ solid electrolytes have even lower values.[43]

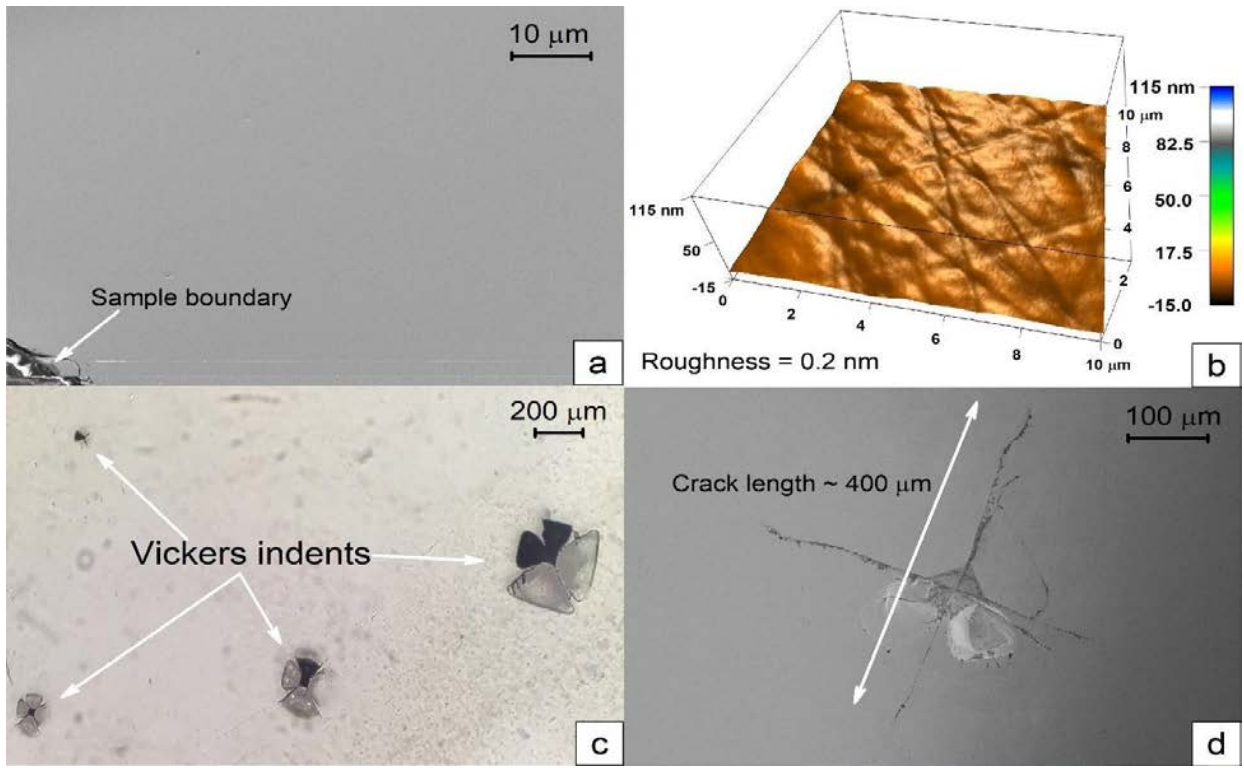

**Figure 2.** (a) SEM image of sc-LLZTO surface polished to a final grit size of 1μm. (b) AFM scan of the sc-LLZTO surface showing linear features from polishing. The rms

roughness is 0.2 nm when measured with a scan size of 10 μm x 10 μm and a tip radius of 7 nm. (c) Optical microscopy image of four Vickers indents produced under loads of 2N, 5N, 10N, and 20N. (d) A representative SEM image of the largest indent, for which the crack length is ~400 μm.

Following indentation, gold electrodes of 2 x 2 mm square or 3 mm diameter circular shape were sputtered onto the sc-LLZTO, being sure to cover the area containing the indents. This gold electrode is the surface onto which lithium metal is subsequently electrodeposited. Because gold readily alloys with lithium metal, the nucleation overpotential required to plate lithium metal at the inception of the experiment is low, [44] ensuring that the applied overpotential is largely transmitted to the electrodeposited lithium metal. A piece of lithium foil (0.75mm thickness, 155 mAh/cm$^2$) was then placed between the solid electrolyte and a stainless-steel current collector to complete the cell assembly. The shape and area of the lithium counter electrode is therefore equal to that of the sc-LLZTO solid electrolyte, which is 20 - 30 mm$^2$ for the samples used. In this configuration, the lithium flux from the metal counterelectrode is focused on the smaller area of the gold working electrode (4 or 7 mm$^2$).

Before electrochemical cycling, the cell was heated to 170 °C for 1h and slowly cooled to room temperature. We found that this procedure greatly reduced the interfacial impedance of the cell, allowing the application of high current densities (up to 10 mA·cm$^{-2}$), while maintaining WE potentials well below 5.0 V. This may be important for avoid SSE oxidation at the positive electrode during the experiment. [45]

## 3. Results and Discussion

### 3.1 *Operando* chronoamperometric experiments

A total of five electrodeposition cells prepared in the above manner were studied, at two different current densities, 10 mA·cm$^{-2}$ (3 samples) and 5 mA·cm$^{-2}$ (2 samples) the current density being defined with respect to the area of the gold electrode. We choose these current densities assuming that a practical lithium metal battery will require an electrode areal capacity of ~5 mAh·cm$^{-2}$ and will need to be charged at rates of 1C-2C. The current densities used correspond to these conditions. A summary of the results is presented in **Table 1**.

| Sample | Current Density (mA/cm$^2$) | Electrode Shape | Indents and location | Short Circuit Observed? | Experiment Duration (s) | Shorting Location | Li volume deposited (10$^{-5}$ cm$^{-3}$) |
|---|---|---|---|---|---|---|---|
| 1 | 10 | 0.2cm x 0.2cm | 4, at center | Yes | 37 | 2, at electrode edge | 2.2 |
| 2 | 10 | 0.2cm x 0.2cm | 4, at center | Yes | 68 | Multiple locations (see text) | 3.7 |
| 3 | 10 | 0.3cm diameter round | 4, at center | Yes | 50 | 3, at electrode edge | 2.2 |
| 4 | 5 | 0.2cm x 0.2cm | None | No | 240 | No shorting | 6.5 |
| 5 | 5 | 0.2cm x 0.2cm | None | Yes | 300 | 1, at electrode corner | 8.1 |

**Table 1.** Summary of results and observations for galvanostatic electrodeposition experiments performed on five Au/sc-LLZTO/Li cells in this study.

Four out of the five samples tested exhibited a short circuit. In each case, lithium metal penetration through the transparent sc-LLZTO was clearly observed. Of these four, 3 were tested at 10 mA·cm$^-$

$^2$ and one at 5 mA·cm$^{-2}$. However, all short-circuited samples exhibited lithium penetration originating at the perimeter of the gold working electrode, away from the introduced defects. One sample exhibited penetration originating from both the edge, the largest introduced indent, and a location in the middle of the electrolyte free of introduced defects.

At 10 mA·cm$^{-2}$ current density, a short circuit was observed in less than 1.5 min; the specific experiments correspond to 2.2-3.7 cm$^3$ of electrodeposited lithium metal. The sample that short-circuited at 5 mA·cm$^{-2}$ did so in 5 min, after which a larger lithium metal volume of 8.1 x 10$^{-5}$ cm$^3$ had been deposited. This indicates that at the lower current density of 5 mA·cm$^{-2}$, more of the deposited lithium is not contributing to the filament but being deposited elsewhere. Below, we discuss the data from two of the cells tested at 10 mA·cm$^{-2}$.

*In-situ* optical microscopy images of **Sample 1** before and after short-circuiting are shown in **Figure 3(a)** and **3(b)**, respectively. (A portion of the sputtered gold electrode is missing at the left side, having been scraped off by the tungsten current collector during setup.) Lithium metal filaments were observed to initiate and propagate within the first 5 seconds after the application of current, as can be seen in the real-time **Video S1** in Supplementary Materials. **Figure 3(c)** shows the voltage-time trace for this experiment, where the sudden drop to zero voltage at 37s corresponds to the growth of a metal filled crack completely through the 2 mm thickness of the single crystal. One filament in particular grows most rapidly in Figure 3(b) to penetration. The elapsed time to short-circuit for this filament corresponds to a plated lithium volume of ~2.2 x 10$^{-5}$ cm$^3$ (~4.5 µAh of charge passed). Notice in Figure 3(b) that both instances of lithium filament formation occur at the perimeter of the gold electrode, and that lithium penetration did not ensue from the large Vickers indents towards the center of the gold electrode.

These observations were consistent amongst the three electrodeposition cells exhibiting a short circuit at 10 mA·cm$^{-2}$: lithium metal penetration initiated at the perimeter of the gold current collector and grew through the single crystal to penetration. *In-situ* optical microscope images of the second electrodeposition cell, **Sample 2**, prior to the experiment and post short-circuit are shown in **Figure S1(a)** and **S1(b)**, respectively. In this second cell, the cell voltage remained relatively steady at 3 V until a short circuit occurred 68 seconds into the experiment, corresponding to a plated lithium volume of ~3.7 x 10$^{-5}$ cm$^3$ (charge passed was ~7.6 µAh). In the third cell that underwent a short circuit, metal penetration into the bulk initiated 5 seconds into the experiment, and lasted for approximately 50 seconds, corresponding to a plated lithium volume of ~2.2 x 10$^{-5}$ cm$^3$ (4.5 µAh of charge passed).

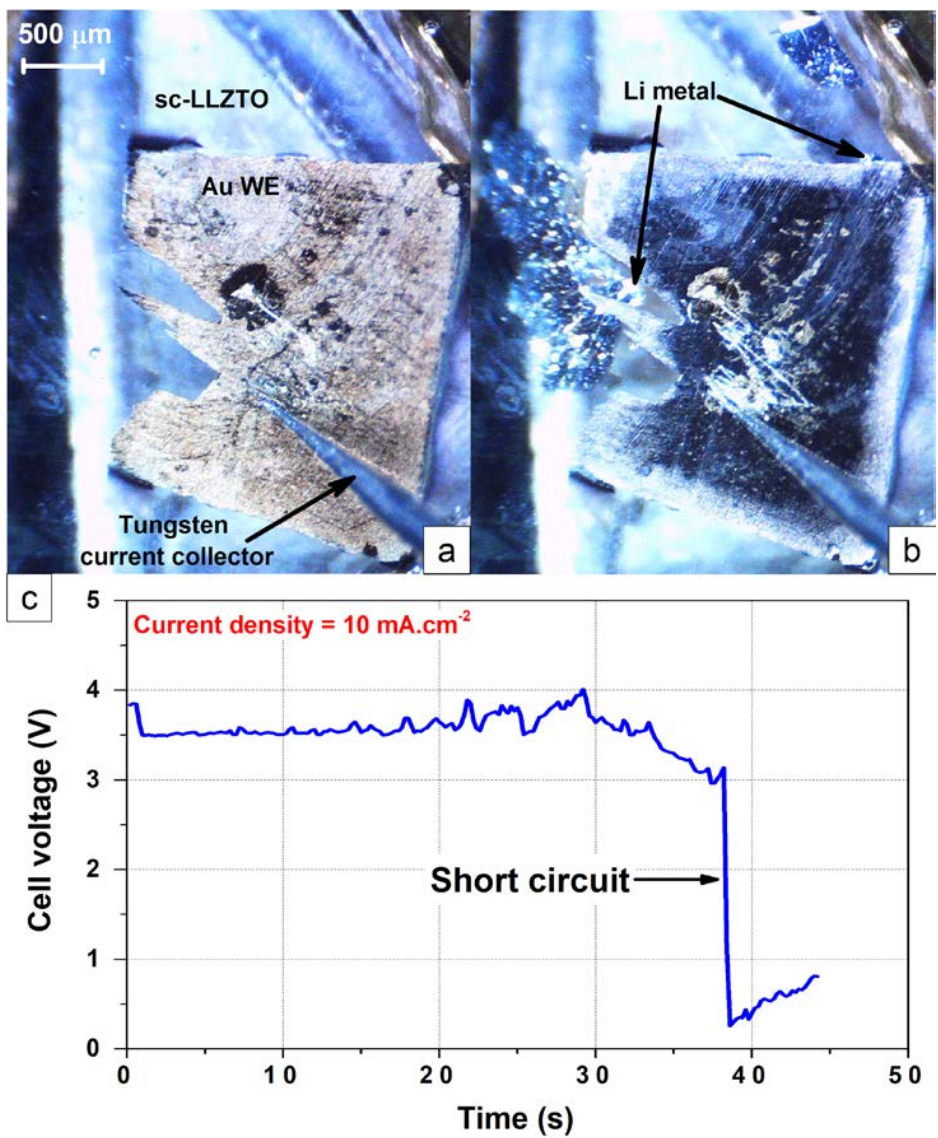

**Figure 3.** *In-situ* optical microscopy images (a) prior to the experiment, and (b) post short-circuit, of a sc-LLZTO during galvanostatic deposition of Li metal beneath the gold WE at 10 mA·cm$^{-2}$ current density. Lithium filaments propagated through the SSE until short circuit occurred 37 seconds into the experiment. (c) The corresponding voltage-time trace, showing a sudden drop upon short-circuiting.

In one of the samples tested at 5 mA·cm$^{-2}$, a short circuit did not initiate within the experimental time of ~240s, which corresponds to a plated lithium volume of ~6.5 x 10$^{-5}$ cm$^3$ (~13.3 µAh charge passed). The experiment was terminated when the cell voltage rose sharply. We believe that this impedance rise corresponds to the formation of a gap at the counter electrode, as the interface between the counter electrode and the sc-LLZTO could be observed to discolor *in situ*. The sample tested at 5 mA·cm$^{-2}$ that did short-circuit did so at ~300 seconds, corresponding to a plated lithium volume ~8.1 x 10$^6$ μm$^3$ (~16.6 μAh charge passed).

The repeated observation that lithium metal penetration initiates at the working electrode edge suggests that current focusing at current collector discontinuities is important. We address this topic in detail in Section 3.3.

**3.2 Subsurface observations**

*Ex-situ* optical microscopy was performed on the sc-LLZTO samples that exhibited a short circuit to examine the initiation sites at which lithium penetrated the single crystal, and the morphology of the lithium filaments within the solid electrolyte. Three key observations are noted. Firstly, crack growth ahead of the lithium filament that did not lead to catastrophic fracture was observed within the electrolyte. That is, cracks that are not filled with lithium metal, as well as those that are, appear to have propagated stably under the electrochemically-created load. Second, plan view observations of the sc-LLZTO after polishing off the gold electrode revealed metal deposition at the introduced Vickers indent in only one instance out of a total of 16 indents. Clearly, the Vickers indents did not dominate the shorting behavior. Thirdly, transverse views of polished single crystals that exhibited a short circuit showed conclusively that the lithium metal penetrated through the entirety of the solid electrolyte.

An optical microscopy image of a lithium filament in Sample 2, taken before the gold electrode was polished away, is shown in **Figure 4(a)**. Here, we focus on a subsurface crack that has formed ahead of a lithium filament. A second image of this site is shown in **Figure S1(c)** in which the aforementioned crack is invisible at a different focal depth**.** These images correspond to the site labelled "Li penetration" in **Figure S1(b)**. Upon imaging through-focus, it was clear that lithium filled cracks are always opaque, and that the crack ahead of the lithium is not filled. This observation shows that crack-opening stresses produced by lithium metal deposition inside of a flaw can produce stable crack extension without catastrophic fracture of the electrolyte. Based on the mechanism proposed in our previous publication, the extension of a lithium filled filament will instantaneously decrease the mechanical driving force (i.e., the strain energy release rate), such that stable crack growth is expected to occur. This is consistent with the observed behavior. However, our original explanation (**Fig. 9** in Porz *et al.*) [23] considers the extension of lithium filled filaments. The new images showing cracks ahead of the filament that do not initially contain lithium indicate that another related phenomenon is possible. Here, it appears that the fracture resistance for the extension of an unfilled crack is lower than that for one filled with lithium. This, for example, is expected to occur if the Li/LLZTO interfacial energy is larger than the LLZTO surface energy. In these cases, the continuing flux of lithium should fill the empty crack tip, ultimately increasing the internal pressure and inducing additional crack extension.

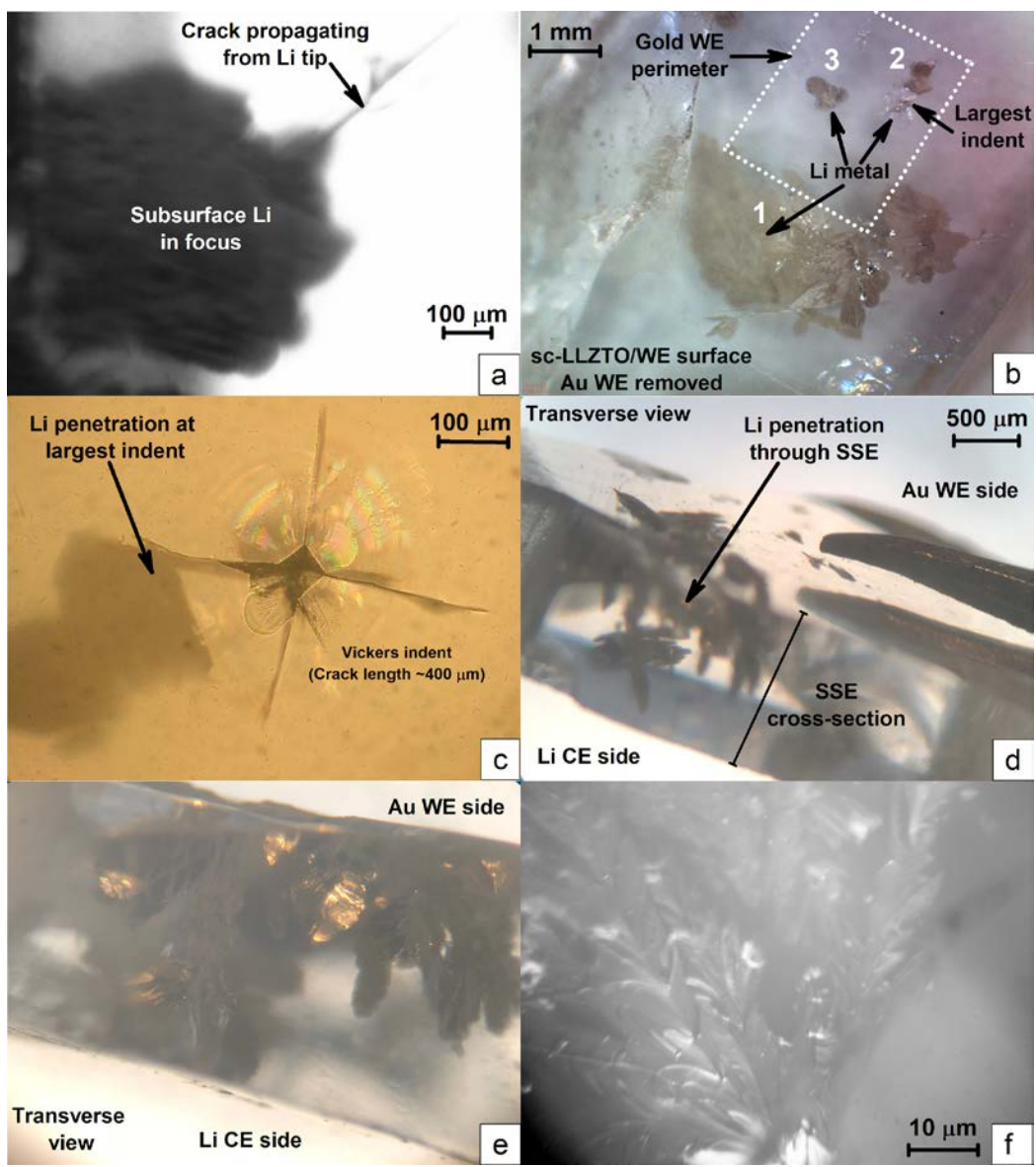

**Figure 4.** (a) Post short-circuit *ex-situ* optical microscopy image of sample 2, focused on a subsurface lithium filament within the single crystal LLZTO. A crack that is apparently free of lithium metal emanates from the lithium filament tip. (b) Optical microscopy image of sample 2 after the gold WE has been removed. The original WE perimeter is outlined in white. Lithium nucleation and growth has occurred at three locations, labeled 1-3. See text for discussion. (c) Magnified view of location 2, showing a Vickers indent and associated corner-cracks, from one of which a lithium metal filament has grown. This is the largest indent in this sample. (d) and (e) Transverse views of the solid electrolyte showing the leaf-

like morphology of the lithium metal filaments that have penetrated into, or completely through, the sc-LLZTO. (f) Magnified optical microscopy image of the leaf-like growth morphology of a lithium metal-filled crack..

An optical microscopy image of **Sample 2** taken after polishing the gold working electrode from the short-circuited sample is shown in **Figure 4(b)**. This image reveals the initiation sites for lithium penetration at the sc-LLZTO surface beneath the gold WE, the perimeter of which is outlined in white. Here we see three lithium filament nucleation regions, labeled 1-3. By far, the greatest amount of lithium metal penetration has occurred from location 1 at the edge of the gold working electrode. This condition was observed in all four of the electrodeposition cells that exhibited a short circuit.

Closer examination of location 2, **Figure 4(c)**, revealed that a lithium-metal filled crack had in fact initiated from one of the corner-cracks produced by the largest of the four Vickers indents. Although this one example conforms to our initial expectations that a large intentional defect would seed the growth of lithium metal, none of the other of the 12 Vickers indents in samples 1-3 showed this behavior. Even more surprising was the observation at location 3 in Sample 2 (Fig. 4b) of a lithium metal filament initiating from a region with no apparent surface defect, yet in close proximity to the intentional Vickers indents. Clearly, other factors such as local electric field intensification are dominating the cell shorting behavior.

Transverse views of Sample 2 taken after short-circuiting are shown in **Figures 4(d)-(e).** Additional plan and transverse views appear in **Figure S2(a)** and **(b)**. The transverse views show the extent of penetration of various lithium filaments from the gold WE. It is firstly clear that some of lithium filaments have penetrated completely through the sc-LLZTO and connect the WE to the CE. The single crystal has not, however, fractured completely despite the internal deposition of lithium metal (occupying a volume of ~2.6 x $10^{-5}$ cm$^3$). Note that what we have described as lithium "filaments" each consists of multiple, planar, metal-filled cracks. The orientation of these crack planes does not appear to follow any particular low-index orientation of the crystal. Single crystal XRD showed that the surface on which the WE is deposited is a {211} plane, and that the polished facet through which the transverse views were taken is {231}. However, LLZTO does not exhibit easy cleavage along a favored plane; this is evident in observations of the single crystal fracture surfaces. Thus, the orientation of the metal-filled crack planes appears to be randomly selected from a crystallographic perspective. Nonetheless, the filaments do not appear to traverse a straight-line path from the WE to the CE, but "fan out" in lateral directions. This is consistent with growth following the electric field distribution, discussed later. A rotating view of **Sample 2** can be observed in real time in **Video S2**.

**Figure 4(f)** shows the morphology of the lithium grown through the solid electrolyte. We note the similarities in morphology between the structure of this filament grown within a single crystal solid electrolyte and the structure of lithium "dendrites" grown through liquid electrolytes. [46-47] A careful investigation into the physical reasons behind this similarity is a focus of ongoing investigations.

**3.3 Field and current intensification at discontinuities in metal current collectors**

Based on the experimental observations, we propose that the preferential initiation of lithium filaments at the working electrode perimeter, followed by penetration through the single crystal electrolyte, is due to electric field intensification at the periphery of the gold working electrode. Finite element modelling was performed to quantify the spatial distribution of this intensification. In high temperature sodium-sulfur batteries where the metal is molten, Virkar *et al*. [29] have previously proposed that electric field intensification can produce short-circuits through ceramic electrolytes.

We model the system as a set of stacked concentric disks representing the gold working electrode, sc-LLZTO SSE, and lithium counter electrode, with a potential difference between the two electrodes. The diameter of the electrolyte disk is significantly larger than both the gold and lithium electrodes, to allow electrode edge effects to manifest. In this configuration the two metal electrodes possess electric field lines that extend into the dielectric layer (the SSE), beyond the perimeter of the metal electrodes. The electric field distribution is the solution to the Laplace equation under Dirichlet (fixed potential) boundary conditions. The solid electrolyte was treated as an ohmic conductor, and numerical solutions giving the electric field distribution between the lithium and gold electrodes were calculated as a function of two dimensionless geometric parameters. $A$ is the WE/CE diameter ratio, and $B$ is the SSE thickness/CE radius ratio. The results in terms of the current density distributions for four values of $A$ (with $B = 1$ throughout) are plotted in **Figures 5(a)-(d)**. For each value of $A$, the transverse view of the cell is shown at the top, and the plan view at the bottom. **Figure 5(a)**, in which $A = 0.5$, closely matches the experimental conditions for the cells in this study. It can be seen that the lithium electrodeposition current density is non-uniform and concentrates at the perimeter when the collector surface area is less than that of the separator and the lithium source. At a position that is 10 microns from the edge, the electric field is a factor of ~5 times higher, as shown in **Figure 6(a)**. In a single-ion conductor such as LLZO (i.e., unit transference number), the amount of lithium metal deposited is exactly proportional to the charge passed. Thus, the repeated appearance of the penetrating lithium filaments at the edges of the current collectors (**Figure 3(a), (b) and Figure 4(b)**, is consistent with a higher current density along the perimeter. Furthermore, the observation that metal-filled cracks propagate from isolated locations, rather than uniformly along the perimeter, is evidence for initiation at metal-filled flaws where the electric field is highest.

The *operando* observations clearly show that the cracks initiate at the interface between the gold electrode and solid electrolyte, as was the case in the previous study using a point probe.[23] It is possible that even an atomically smooth, defect-free interface will succumb to local non-uniformities in lithium plating, leading to the propagation of metal-filled cracks at sufficiently high driving force. However, in the present case the direct observations of the sample surface (Fig. 2a, b) show that polishing leaves behind features of tens to hundreds of nanometer scale. Pre-existing defects of this size should require local overpotentials of a fraction of a volt to propagate,[23] and given that the present galvanostatic tests produce cell voltages of 3.5 ~ 4V, it is highly likely that the population of surface defects contains ones of supercritical size. We believe that the intersection of the intensified electric field with such defects leads to the growth of the lithium metal filaments, to the exclusion of the larger, intentionally introduced Vickers defects.

We can further use the simulations to understand the results for the current experimental geometry, to extrapolate from those effects to large-area planar battery cells, and to model the effects of a different kind of defect - namely, discontinuities in electrodes and current collectors that may cause local field amplification. Considering first the geometric variable $A$, which represents the ratio of positive electrode to negative electrode area, the simulations show that the current density amplification factor (i.e., obtained 10 micrometers away from the WE perimeter) decreases as $A$ increases; for $A = 0.5$, 1.0 1.5 and 2, the amplification is respectively ~5, ~4.5, ~2 and ~1 (**Figure 6(a)**). We view these as lower bound estimates of the amplification near the edge, since the solution for the electric field diverges exactly at the edge of the electrodeposition electrode when $A = 0.5$, 1.0 1.5 and 2. This lack of convergence is associated with a mathematical singularity in the electric field at the electrode boundary. We somewhat arbitrarily cite the value 10 micrometers away, based on the premise that this region will contain a large number of sub-micrometer defects like those observed in **Figure 2**.

While in lithium-ion batteries the negative current collector usually slightly overlaps the positive current collector, the area ratio $A$ is nonetheless close to one. Our results show that it may be advantageous to design such that $A>1$ to diminish the field amplification. Notice that in **Figure 6**, field enhancement at the edges is completely avoided when $A=3$. While the critical value of $A$ at which edge field enhancement vanishes will vary with geometry, in general, one of our key findings is that a larger positive electrode will help to mitigate electrode edge failures during the charging process of interest to lithium metal batteries.

The second geometric variable $B$ represents the thickness-to-width aspect ratio of the battery, and for a typical thin-plate design will be less than ~0.1. Upon holding $A$ constant and decreasing $B$, the field amplification 10 microns from the current collector edge decreases, but the singularity at the edge is not removed, if $A \leq 1$ and the electrolyte has greater width than the electrodes. Only in the limiting case where the electrolyte width is exactly equal to the width of the negative electrode (and $A \leq 1$) does the singularity vanish. Furthermore, as $B$ decreases, the oversizing of the positive electrode at which the edge field enhancement vanishes also decreases. (This occurs between $A = 1$ and 1.5 for $B = 0.2$, and between $A = 2$ and 3 for $B = 1$). These trends are shown in **Figures 6a) and b)**, respectively. Although the ratio $B$ for a practical battery is likely to be far below 0.2, examination of smaller values of $B$ did not significantly affect the dependence of field amplification on $A$ and electrolyte width as stated above. Therefore, in general, battery designs utilizing a smaller value of $B$ will minimize the positive electrode oversizing required to mitigate electrode edge failures.

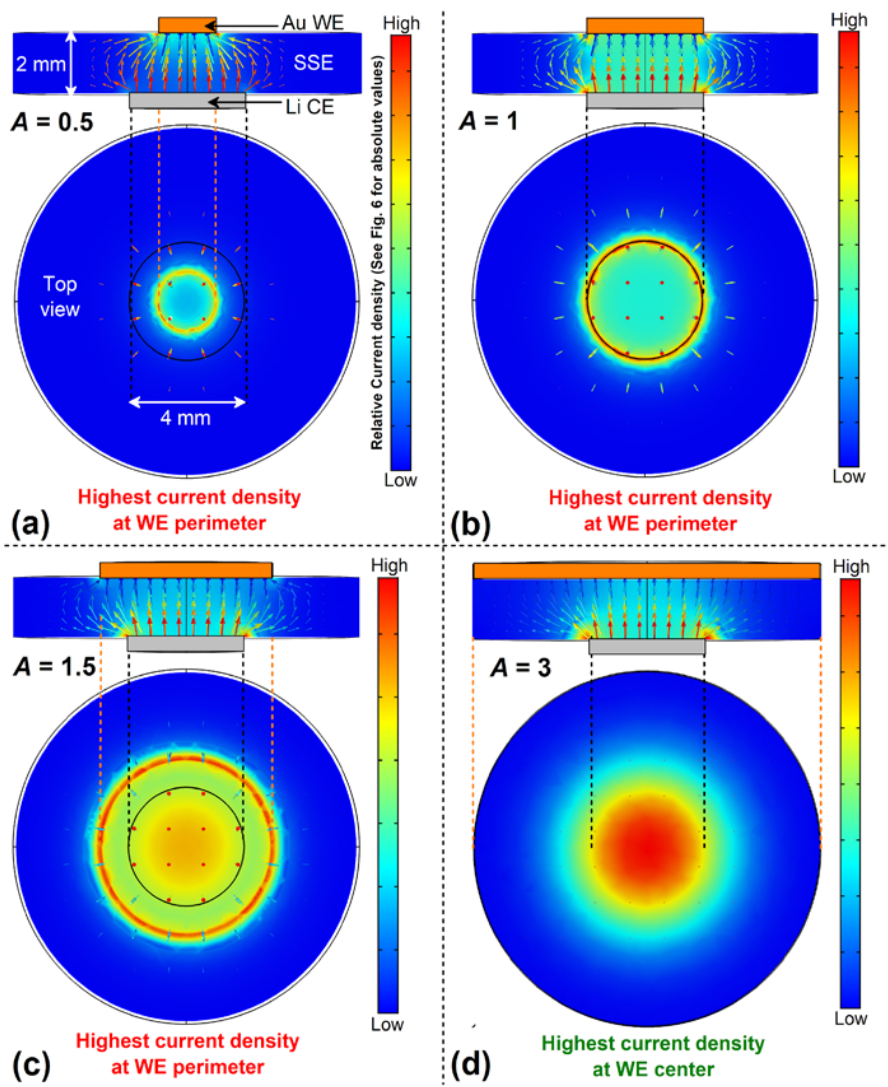

**Figure 5.** Current density contour plots for an Au/SSE/Li cell with a WE/CE diameter ratio (***A***) of (a) 0.5, (b) 1, (c) 1.5, and (d) 3. Corresponding values for electric field are shown in **Figure 6**. The SSE and Li CE diameters are fixed at 4 and 12 mm respectively, and the SSE thickness is fixed at 2 mm. (Accordingly, ***B***, the SSE thickness/CE radius ratio, is one). ***A*** = 0.5 closely represents the experimental conditions in this study. For ***A*** = 0.5, 1, and 1.5, the highest current densities are observed at the WE perimeter, whereas between ***A*** values of 1.5 to 3, the current density intensification transitions the center of the electrode.

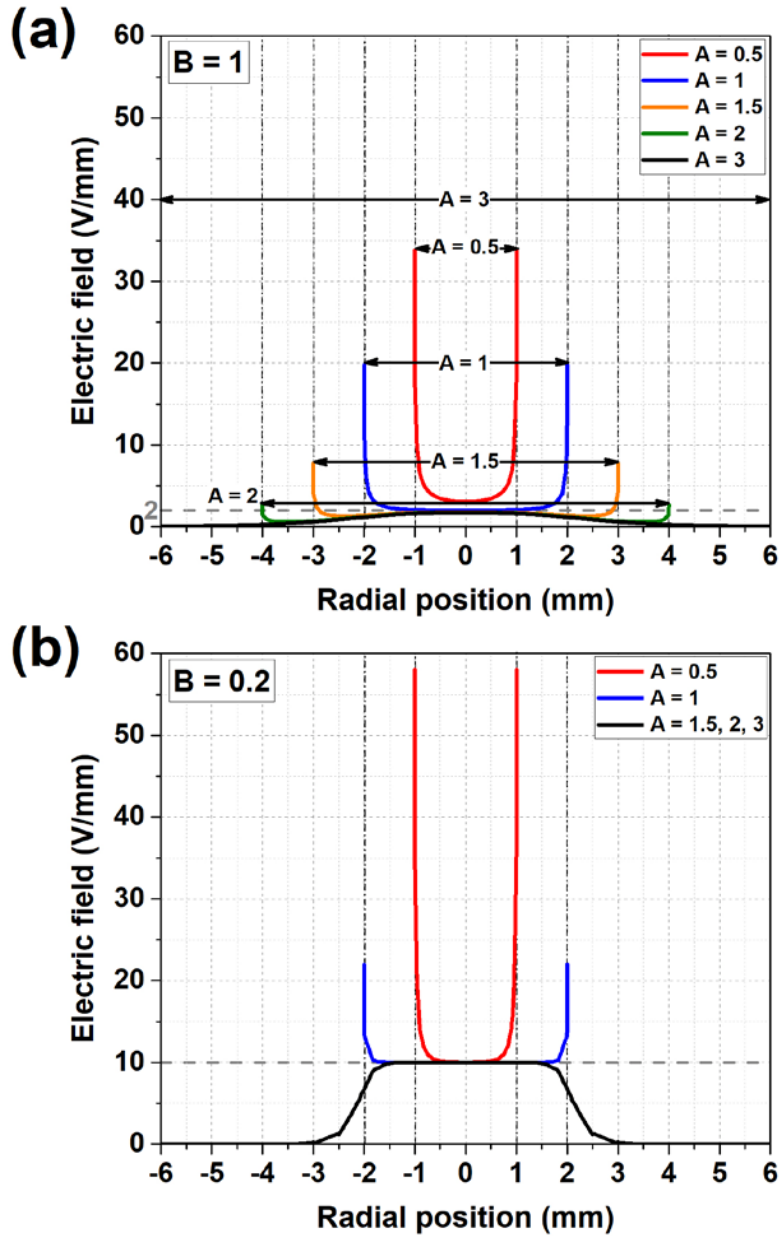

**Figure 6.** The spatial distribution of electric field for WE/CE diameter ratios (***A***) of [0.5, 1.0, 1.5, 2.0, and 3.0] for electrolyte thickness/CE radius ratio (a) (***B***) = 2 and (b) ***B*** = 0.2. The light grey line represents the macroscopic value of the electric field, defined as the potential difference divided by the electrolyte thickness. SSE and lithium CE diameters are 4 and 12 mm respectively, and SSE thickness is 2 mm. $B = 1$ approximates the experimental conditions in this study.

## 4 Conclusions

Lithium electrodeposition experiments were performed using single-crystal $Li_6La_3ZrTaO_{12}$ garnet as a model brittle solid electrolyte to investigate the factors that determine lithium metal penetration. *In situ* and *ex situ* microscopy was performed on cells undergoing galvanostatic electrodeposition at realistically high current densities for practical applications. Lithium infiltration resulting in short circuits occurred on the minute time scale at current density of 10 mA/cm$^2$. The initiation of lithium metal filled cracks occurred predominantly at the perimeter of the working electrode, even when much larger (up to 0.4 mm), intentionally produced surface defects are present nearby. After crack initiation, subsurface crack extension is observed that is consistent with stable crack growth, i.e., without catastrophic fracture.

The crack initiation sites coincide with the locations of maximum electric field at current collector edges. It is concluded that electric field amplification drives lithium penetration from sites on the solid electrolyte surface with supercritical surface flaws. Finite element modelling was used to investigate the magnitude of field enhancement as a function of the relative sizes of the positive and negative electrodes, and the thickness-to-width aspect ratio of the cell. It is concluded that the risk of edge effects resulting in lithium metal penetration (i.e., during charging of a lithium metal battery) is reduced by designing cells to have a larger positive electrode than negative electrode, and a smaller thickness/width ratio.

## 5. Experimental methods

*Synthesis*: The single crystal LLZTO sample was grown by the Czochralski method directly from the melt. The starting materials, $Li_2CO_3$ (99 %, Merck), $La_2O_3$ (99.99 %, Aldrich), $ZrO_2$ (99.0 %, Aldrich), $Ta_2O_5$ (99.99 %, Aldrich) were dried, mixed in a stoichiometric ratio with 10 wt% excess of $Li_2CO_3$, then pressed and sintered at 850ºC for 4 h at a heating rate of 5ºC/min. The pellet was then melted in an iridium crucible by RF-induction heating using a 25 kW microwave generator. An iridium rod was used as a seed for crystal growth under a nitrogen gas atmosphere. 1.5 mm/h and 10 rpm were the seed pulling and rotation rates, respectively. The transparent crystal was then cut into samples of approximately 5 x 5 x 2 mm dimensions.

*Sample preparation and characterization:* LLZTO single crystals were embedded in epoxy, then polished using an EcoMet™ 250 Pro Grinder Polisher (Buehler, Lake Bluff, Illinois, USA). The samples were first polished using 600 grit size SiC abrasive paper to reveal two faces. To obtain finely polished faces, the samples were then sequentially polished using aqueous diamond suspensions of 9 μm and 1 μm particle size, for 25 min and 10 min, respectively. The samples were then sonicated for 5 min in deionized water to remove polishing debris. AFM surface scans of the sample were taken with an Asylum Research Cypher AFM (Asylum Research, Santa Barbara, California, USA). SEM imaging of the samples was conducted using a Merlin GEMINI II SEM (Carl Zeiss Microscopy, Jena, Germany) operating at 15 kV accelerating potential and 215 pA current. All samples were prepared in an argon-filled glovebox (Oxygen and Water levels below 0.1 ppm) and transported to the SEM using an anaerobic transfer box, the design and operation of which is described elsewhere. [48] A LECO LM248AT Microindentation hardness testing system was used to produce the controlled Vickers indents on the sc-LLZTO surface. The crystallographic orientation of the polished sc-LLZTO facet parallel to the working and counter electrodes, the crystal facet representing the cross-sectional surface, and the direction of the Li filaments internal to the LLZO crystal was performed with a Bruker D8 Discover X-ray diffractometer using a CuKα X-ray source and a Vantec 2000 area detector.

*Experimental setup*: Galvanostatic experiments were conducted on the samples using a Solartron 1400/1470E cell test system. *In-situ* lithium plating activity unto the working Au electrode was recorded using an optical microscope (Firefly Global GT825, Belmont, Massachusetts, USA). *Ex-situ* microscopy was conducted using the Olympus BH (Olympus, Shinjuku-ku, Tokyo, Japan) and Stereomaster II SPT-ITH (Fisher Scientific, Hampton, New Hampshire, USA) microscopes.

*Finite element modelling*: Gold WE, lithium CE, and SSE diameters were initially set to 2, 4, and 12 mm, respectively (WE/CE diameter ratio = 0.5). The SSE thickness was set to 2 mm, and the voltage across the SSE was set to 4 V. These parameters closely matched experimental conditions. To our knowledge, the dielectric constant of LLZO garnets have not been measured; here the value was set to 3.8, the dielectric constant of quartz (a representative ceramic material). The assumed value of the dielectric constant does not affect the current and electric field intensification factors of interest here. The spatial distribution of the electric field and hence the current density (being linearly proportional to electric field) was computed using the electrostatics module of the COMSOL Multiphysics finite element software.

**Conflicts of Interest**

There are no conflicts to declare.

**Acknowledgements**

Tushar Swamy and Richard Park contributed equally to this work. The authors acknowledge support from the US Department of Energy, Office of Basic Energy Science, through award number DE-SC0002633 (J. Vetrano, Program Manager). BWS acknowledges support by the US Department of Energy, Office of Basic Energy Science under Contract DE-SC0018113. DR acknowledges the Austrian Science Fund (P 31437–N36). The authors also acknowledge use of the MIT Nanomechanical Technology Laboratory (A. Schwartzman, Manager).